\documentclass[reprint,nofootinbib]{revtex4-1}

\usepackage{graphicx}
\usepackage{amssymb}
\usepackage{amsmath}
\usepackage{epsfig}
\usepackage{chemarr}
\usepackage{url}
\usepackage{natbib}
\usepackage{dcolumn}
\usepackage{bm}
\usepackage{color}

\begin{document}

\title{Multifarious Assembly Mixtures: Systems Allowing Retrieval of Diverse Stored Structures}

\author{Arvind Murugan*}
\affiliation{Kavli Institute for Bionano Science and Technology and the School of Engineering and Applied Sciences, Harvard University, Cambridge, Massachusetts 02138}
\author{Zorana Zeravcic*}
\affiliation{Kavli Institute for Bionano Science and Technology and the School of Engineering and Applied Sciences, Harvard University, Cambridge, Massachusetts 02138}
\author{Michael P. Brenner}
\affiliation{Kavli Institute for Bionano Science and Technology and the School of Engineering and Applied Sciences, Harvard University, Cambridge, Massachusetts 02138}
\author{Stanislas Leibler}
\affiliation{Laboratory of Living Matter, The Rockefeller University, 1230 York Avenue, New York, NY 10065}
\affiliation{Simons Center for Systems Biology, School of Natural Sciences, Institute for Advanced Study, Princeton, NJ 08540}


\thanks{These authors contributed equally to this work.}


%

\begin{abstract}
Self-assembly materials are traditionally designed so that molecular or meso-scale components form a single kind of large structure. Here, we propose a scheme to create ``multifarious assembly mixtures", which self-assemble many different large structures from a set of shared components. We show that the number of multifarious structures stored in the solution of components increases rapidly with the number of different types of components. Yet, each stored structure can be retrieved by tuning only a few parameters, the number of which is only weakly dependent on the size of the assembled structure. Implications for artificial and biological self-assembly are discussed.
\end{abstract}


\maketitle

\section*{Introduction}
{A} classical example of self-assembly is crystallization. At low temperatures the crystalline phase is typically stable, and thus grows spontaneously from solution through homogeneous nucleation.  If several competing crystalline phases are allowed by microscopic interactions, the efficient production of a desired phase often requires heterogeneous nucleation from a seed of this phase, with precise annealing protocols \cite{Lifshitz196135,binder76}. More complex microscopic interactions may lead to a glassy phase with many competing structures; however, it is generally impossible to control local compositions or microscopic interactions to obtain a particular structure.

Recently, there has been a dramatic change in macromolecular and colloidal assembly techniques, made possible by the use of biopolymers, such as DNA, to create a large variety of inter-component interactions. Indeed, biomolecules offer exquisite control of microscopic interactions that allows self-assembly of diverse large structures. Examples range from nanoparticle assemblies  \cite{Mirkin:1996th,Alivisatos:1996vs,Valignat:2005p8423,Biancaniello:2005p5460}, which can also form macroscopic crystals \cite{Park:2008ur,Nykypanchuk:2008cp,Kim:2006p8449,Macfarlane:2011p18077}, to structures using DNA itself as a building material. In the latter case, DNA origami uses short DNA strands to controllably fold a long backbone strand into different well controlled structures \cite{rothemund2006folding,douglas2009self}, while short strands of DNA by themselves can also build up complex three dimensional objects \cite{Wei:2012ip,Ke:2012jd}. Similar efforts are underway using rationally-designed proteins by creating complementary binding sites on their surfaces \cite{King:2012gc,Lai:2012vh,King:2014cf}. Up until now, however, experimental and theoretical studies have been limited to devising interactions for the assembling a single structure.  This is to be contrasted with biological systems, where many different self-assembled structures can be formed within the cell cytoplasm. These assembled structures can in fact share some of their components and can be dynamically induced independently from one another \cite{Kuhner27112009}.

Here, we propose a new mechanism for the self-assembly of many different structures from one large set of shared components.  Each structure is multifarious, i.e., is made out of many different types of components. Such self-assembling systems, which we propose calling ``multifarious assembly mixtures'', are stable and yet responsive. This means that the mixtures do not form structures spontaneously, but can be controllably induced to assemble a specific structure. Different structures are ``encoded"  through the choice of molecular interactions and thus ``stored" in the mixture, to then be ``retrieved"  by changing only a small number of parameters.

The theoretical framework introduced below allows calculation of the capacity of these systems, i.e., how many different independent structures can be stored and retrieved in a mixture of $N$ species of components. In the traditional approach of self assembly without shared components, if each structure S is composed of the same number $N_S$ of different species, only $N/N_S$ different structures can be self-assembled. In contrast, in multifarious assembly mixtures, many more distinct structures can be stored. Any stored structure can be retrieved with a (super)critical nucleation seed. Multiple  seeds can induce the simultaneous assembly of multiple corresponding structures. Moreover, we show that each  different structure can be retrieved by changing only a small number of chemical potentials or interspecies interactions, where the number of tuned components is the number of components in a (super)critical nucleation seed of the desired structure. Classical nucleation theory implies that the size of this seed is only weakly dependent on the size of the structure that is built.

\subsection{Model}
\begin{figure*}[h!]
\centering
\includegraphics[width=0.9\textwidth]{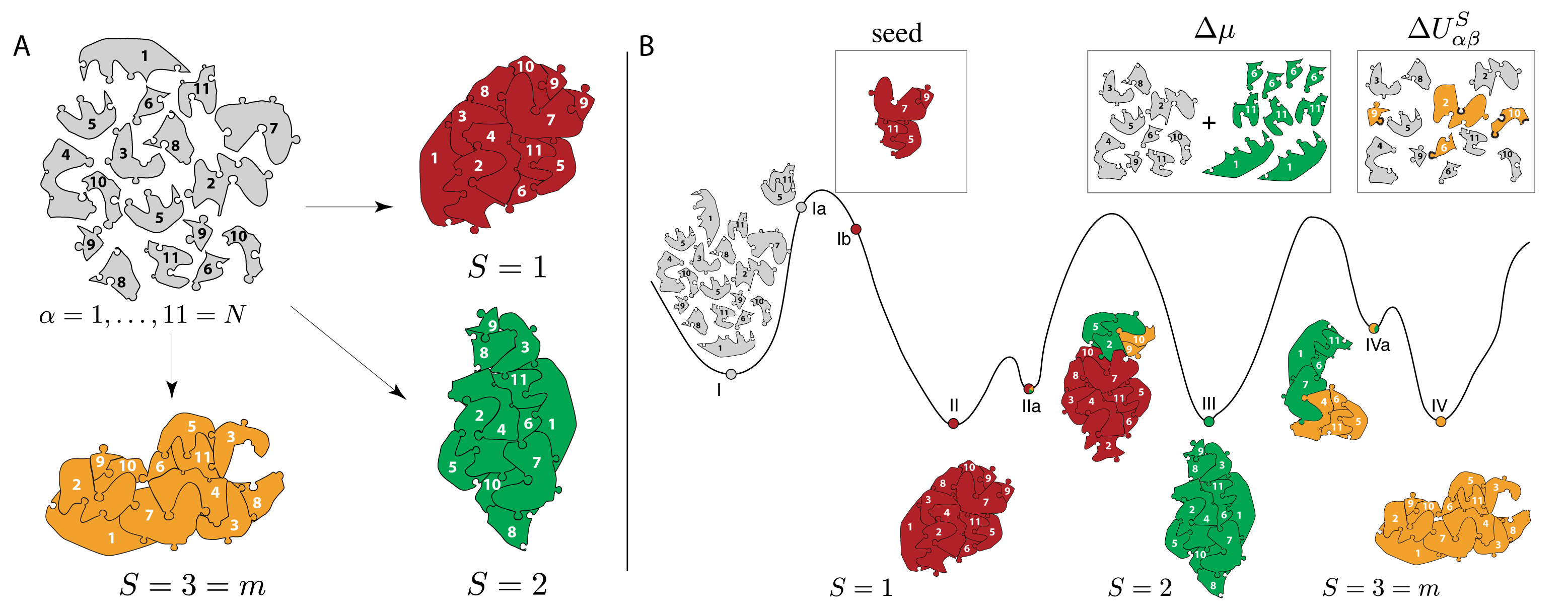}
\caption{A) Schematic depiction of the basic idea of assembly of different desired multifarious structures ($S=1, S=2$ or $S=3=m$) by using the same set of components. In general, the multiplicity of different component species within a structure can be non-trivial, i.e., the number of component species, $N_S$, comprising structure $S$ can be different from the size of that structure, $M_S$, $N_S\neq M_S$. For example, in structure $S=1$, the multiplicity of species $9$ is $n^1_9=2$. Similarly, $n^2_8=2$ and $n^3_3=2$ in structures $S=2$ and $S=3$ respectively. B) Free energy landscape and chimeric states. (I) A solution of $N$ different species of components, with interactions designed for assembly of desired structures $S=1$ (II), $S=2$ (III) and $S=3=m$ (IV). The desired stored structures are not the only free energy minima; chimeric structures, i.e., hybrids between different stored structures, can also exist (IIa, IVa). Insets --- Assembly of the stored structures can be triggered by manipulating a small number of components: (Left) Introducing a supercritical seed, a subcluster of the desired stored structure; (Middle) Increasing the average concentration of components that can make a supercritical seed by tuning their chemical potentials; (Right) Increasing the specific binding energy of components that can make a supercritical seed.}
\label{fig:GeneralIdea}
\end{figure*}

Consider $N$ species of interacting components in a solution kept at a constant temperature $T$. In principle, each species can have a different chemical potential $\mu_{\alpha}$, $(\alpha = 1,...,N)$, but for simplicity we assume for now that the chemical potentials have the same value $\mu$. We would like the components to be able to self-assemble into one of $m$ distinct, multifarious structures, $S=1,\ldots,m$, on demand (Fig.~\ref{fig:GeneralIdea}A).

A typical multifarious structure $S$ is built of $N_S$ component species. In general, each species $\alpha$ in the structure $S$ has its own multiplicity $n^{S}_{\alpha}$. In contrast to traditional studies of self-assembly, e.g. of crystals, where the same component species appears in many copies in the assembled structure, for multifarious assembly mixtures we are interested in the case of small values of $n^{S}_{\alpha}$. Indeed, for simplicity, we assume here that all component species have a single copy in every stored structures, $n^{S}_{\alpha}=1$, so that the number of species $N_S$ used in the structure equals the size of the structure $N_S=M_S$. Additionally, we make a simplifying assumption that all the structure sizes $M_S$ have the same value $M$, so $N_S=M_S=M$.

Both cellular systems and recent DNA-mediated assembly experiments show that a single structure $S$ can be robustly assembled if each pair of neighboring components, of species $\alpha$ and $\beta$ ($\alpha,\beta \in \{1,\ldots, N\}$) interact through a specific binding interaction. Our next simplifying assumption is that all these interaction energies are equal: $U^S_{\alpha\beta} = - E$ and we will also set all non-specific interactions to zero. The binding interactions between different components are mediated through a discrete number of ``binding sites", with a species $\alpha$ having a valence $z_\alpha$. For simplicity we assume that all components have the same valence $z$.

How might we choose an interaction energy matrix $U^{tot}_{\alpha \beta}$ so that the components are capable of assembling different desired structures $S=1,\ldots,m$ (Fig.~\ref{fig:GeneralIdea}A)?  The simplest general prescription that can work for arbitrary structures is to assume that two species $\alpha$ and $\beta$ bind specifically with energy $-E$ if and only if at least one of the desired structures $S$ requires this binding. Such a  matrix $U^{tot}_{\alpha \beta}$ then has the potential for ``storing'' each structure $S$ as a local free energy minimum (Fig.~\ref{fig:GeneralIdea}B). This matrix can be written as,
\begin{eqnarray*}
U^{tot}_{\alpha \beta} = 
	\left\{ 
		\begin{array}{cl}
			 -E & \mbox{\small{if $\alpha$, $\beta$ interact specifically ($U^S_{\alpha \beta} = -E$) in any $S$,}}\\
			    0 & \mbox{\small{otherwise.}}
		\end{array}
	\right.
\label{eqn:pooledU}
\end{eqnarray*}
This form of energy matrix implies that component species can be promiscuous in their interactions. Indeed, since a given species $\alpha$ binds specifically to its partners in each of the stored structures, the total number of specific binding partners for species $\alpha$ can be large. 

In addition to the free energy minima corresponding to the desired structures, other undesired local minima might emerge. These correspond to chimeric structures, or ``chimeras", made of chunks of different desired structures that can bind together due to the promiscuity implied by Eqn~\eqref{eqn:pooledU}. The stability of the stored structures is determined by the size of the free energy barriers between the different minima.  For instance, if the barriers are low, chimeras will form spontaneously, even if their local free energy minima lie higher than those of desired structures (Fig.~\ref{fig:GeneralIdea}A). Similarly, the free energy barriers between the solution of unbound components and other minima determines the solution's characteristic time $t_*$, beyond which stored structures nucleate spontaneously and the process of the controlled retrieval of stored structures is compromised. Thus, $t_*$ is the functional ``lifetime'' of the multifarious assembly mixture.

\subsection{Storage Capacity}
How many different multifarious structures, each of size $M$, can one store by using $N$ different species of components with well-chosen interspecies interactions defined by Eqn~\eqref{eqn:pooledU}? If each species contributed to only a single structure, the maximum capacity would be $N/M$. By sharing species between structures, however, a much larger number of structures can be stored before chimeras start to dominate. To find this increased capacity, consider components attaching to the boundary of a growing seed. The promiscuous interactions implied by Eqn.~\eqref{eqn:pooledU} might allow the seed to bind different sets of components, resulting in chimeras. Therefore, let us compute the number of species that can specifically bind to a given boundary site of the seed. Since each component in the bulk of a stored structure has $z$ nearest neighbors, for an incoming component to bind stably, it must form specific bonds with $z/2$ components on the seed's boundary. Due to the promiscuous nature of Eqn.~\eqref{eqn:pooledU}, each of these $z/2$ boundary components can bind specifically to a set of $\sim m \frac{M}{N}$ other species \footnote{To see this, note that if each structure of size $M$ is randomly constituted from the $N$ species, a given species will occur in $\sim m \frac{M}{N}$ of the $m$ stored structures and typically have a different partner in each of them. Hence, a typical species will have $\sim m \frac{M}{N}$ specifically binding partners.}. 
For randomly constituted $m$ structures, each set contains a fraction $mM/N^2$ of all the $N$ component species. The intersection of these $z/2$ sets, of the size $N (m M/N^2)^{z/2}$, determines the species that can specifically bind to all the $z/2$ boundary components. When this number is larger than $1$, many different species can attach to a given boundary site on a growing seed, resulting in a proliferation of chimeras. Hence, the largest number $m$ of structures that can be stored is
\begin{equation}
\label{eq:capacity}
m_c \sim \biggl(\frac{N}{M}\biggr) N^{(z-2)/z}.
\end{equation}
For $z>2$, the exponent $(z-2)/2$ is positive and this equation implies that the capacity $m_c$ can be much larger than the traditional estimate of the capacity $N/M$. It is instructive to understand why $z=2$ structures i.e., linear chains, cannot share components. Binding to an end of a growing chain requires forming a bond with just one component. If that component is promiscuous, the seed can always grow in a non-unique chimeric manner. Hence the promiscuity of individual species, implied by Eqn~\eqref{eqn:pooledU}, must be countered by the requirement on incoming particles to form multiple (i.e., $z/2$ $\ >1$) bonds.

\subsection{Retrieval}
The above argument shows that the number of structures that can be stored and stabilized with $N$ components is large. For this to be useful, we need to be able to retrieve each of them easily. The retrieval can be done in three different ways. One can introduce a nucleation seed, i.e., a part of a stored structure, into the solution. Alternatively, one can enhance the formation of such a seed by increasing the chemical potential of its components by an appropriate amount $\Delta\mu$, or by strengthening the interactions $U^S_{\alpha\beta}$ by $\Delta U$ for bonds found in such a seed. These methods enhance the nucleation of one stored structure without nucleating others, despite all stored structures being made of the same set of components. Such selective nucleation is possible only for multifarious structures;  it relies on the fact that small contiguous subsets of distinct structures typically have distinct compositions. Such subsets can be used as selective nucleating seeds, or to selectively lower the nucleation barrier for one structure using the other two methods described above.

The critical question is how many different species have to be tuned in this way to successfully retrieve a particular stored structure. The answer follows directly from general nucleation theory, which specifies a critical nucleation radius $r_*$ in terms of the chemical potential $\mu$ and bond energy $E$ \cite{oxtoby1998nucleation}.  The minimal seed size $N_*$ needed to recover a structure is set by $r_*$; smaller seeds dissolve back into components while larger seeds are supercritical and grow into stored structures. We can make the multifarious assembly mixture responsive to smaller seeds by lowering the critical nucleation radius, for example, by lowering the ratio $\mu/E$ of chemical potential to bond energy (see SI text and SI Figs.~ S2, S5 and S6).  

However, lowering the critical nucleation radius also lowers the barrier to spontaneous homogenous nucleation. As noted above, critical seeds can spontaneously assemble on a characteristic timescale $t_*$ and grow into random stored structures, without any external input. Thus, at a minimum, we need $t_*$ to be much longer than the retrieval time, i.e., time necessary for a supercritical seed to grow into a full structure. Nucleation theory, adapted to multifarious $d$-dimensional structures, determines $t_*$ as
\begin{equation}
\log\biggl(\frac{t_*}{\tau}\biggr)=\frac{F_*}{ k_B T} -\log(q(m,M)),
\label{eqnuc}
\end{equation}
where $F_* \sim \gamma r_*^{d-1}$ is the free energy barrier, $\gamma$ is the free energy per area required for creating the critical seed and $\tau$ is a time scale connected with microscopic processes. The second term on the right hand side arises because we must account for the multiplicity $q(m,M)$ of distinct nucleation paths leading to the $m$ different stored structures. For small $m$, we can estimate $q(m,M) \sim m M$ to account for critical seeds from different parts of the $m$ stored structures of size $M$ each (see SI Text).

For a fixed $t_*/\tau$, Eqn.~\eqref{eqnuc} can be solved for the nucleation radius $r_*$,  and hence the minimal number of components $N_*$ that must be tuned to retrieve a structure. If all the components have a typical size $a$, this number $N_*$ is of order
\begin{equation}
N_* \sim \biggl(\frac{r_*}{a}\biggr)^d =\biggl(\frac{k_B T}{a^{d-1} \gamma}\biggr)^{d/(d-1)} Q(m,M,t_*/\tau),
\label{eqn:tradefoff}
\end{equation} 
where $Q$ depends only weakly (logarithmically) on $m$, $M$ and $t_*/\tau$. We thus conclude, that since $N_*$ is determined by the nucleation barrier, it is essentially independent of the size of the structure, $M$, that is being retrieved. Note that the above equations show an unavoidable trade-off: increasing the lifetime of the multifarious assembly mixture $t_*$ necessarily increases the nucleation radius $r_*$ and hence increases $N_*$. Thus a more stable multifarious assembly mixture requires a larger seed for recovering stored structures.  

\subsection{Simple Lattice Model} 
In order to study different regimes of self-assembly of multifarious structures, we have considered assembly based on Eqn.~\eqref{eqn:pooledU}, on a simple $2d$ square lattice. Individual components are square tiles that can be one of $N=400$ species. All $m$ stored structures consist of $M=N$ tiles, each tile being of different species positioned inside a $20 \times 20$ square block. More precisely, we assume that all the species are present in all the structures, and each species appears only once in each structure, $n_{\alpha}^S=1$ for all $\alpha$ and all $S$, so that $N=N_S=M_S=M$ for all structures $S=1,\ldots,m$. In other words, each stored structure is simply a different random permutation of the tiles inside the square block. We assume that each tile component can bind up to $z=4$ neighbors through specific binding interactions given by  Eqn~\eqref{eqn:pooledU}, and that all species of tiles have the same chemical potential $\mu$. We run grand canonical Monte Carlo simulations with different numbers $m$ of stored structures on a square lattice of total size $40 \times 40$, for different values of temperature $T$ and chemical potential $\mu$ (see SI Text and SI Fig.~S3).

Starting from a particular supercritical seed (of linear size $r>r_*$) of one of the $m$ stored structures, Fig.~\ref{fig:PD} shows a diagram of the different outcomes of our simulations, as a function of the number of stored structures, $m$, and the temperature $T$ (or more precisely, $k_BT/E$, where $E$ is the specific binding energy), for a fixed $\mu$. We visualize the different stored structures with different colors, with the desired structure colored in dark red. For low $m$ and $T$, the supercritical seed indeed grows into the desired structure. In this regime of parameter space (regime I), the solution behaves as a useful multifarious assembly mixture: the mixture is stable for a long time $t_*$ and stored structures can be retrieved through heterogeneous nucleation \footnote{Our results show that the simulation dynamics obeys the prediction of classical nucleation theory. For instance, within the recovery regime the timescale for appearance of a supercritical cluster, $t_*$, is much longer than the timescale for recovery, $t_{recovery}$. Thus, even though the Monte Carlo dynamics do not reflect the dynamics of a realistic self-assembly system (see e.g.\cite{Reinhardt:2014wj}), they do substantiate the predictions of nucleation theory, and expose different regimes of self-assembly of multifarious structures.}. For higher number of stored structures $m$ (and at higher temperatures $T$) another behavior appears (regime II). It is characterized by the spontaneous homogeneous nucleation of all stored structures from the solution: in this regime, the multifarious assembly mixture is too short lived to allow the structure retrieval, i.e., $t_*$ becomes comparable to the time taken for a supercritical seed to grow into a full desired structure, $t_{recovery}$. At even higher values of $m$ we find yet another regime of behavior (regime III), where chimeric structures  dominate. Finally, at high temperatures $T$, and for all values of $m$, we encounter regime IV, where any initial seed disintegrates into small clusters of individual components. The extent of different regimes depends of course on the chosen model parameters. In particular, the chemical potential $\mu$  influences the extent of regimes I and II (see SI Text and SI Figs.~S7 and S9).

Simulations presented in Fig.~\ref{fig:diffSA} confirm that, in regime I, the assembly of a structure can be triggered not only with supercritical nucleating seeds, but also by enhancing the chemical potential of a small set of tile species, or by increasing the bond energies between the tile species from such a set.

Numerical simulations are also a way to gauge the capacity of an multifarious assembly mixture to store structures, and to compare it with the theoretical predictions presented above. To do this, we have introduced the entire target structure as a supercritical seed, and have examined it after a fixed simulation time chosen to be shorter than the mixture's lifetime $t_*$. We have assessed the quality of retrieval by measuring the error, i.e., the fraction of the final assembled structure that differs from the initial target structure (see SI Text and SI Fig.~S4). Fig.~\ref{fig:scaling}A depicts the error as a function of the number of stored structures $m$, for different number of particle species $N$ (structure sizes being $M=N$), at fixed temperature $T$ and chemical potential $\mu$. There is a transition at critical value $m=m_c$, above which the error rises rapidly. We show that the error curves for different $N$ collapse onto each other when plotted against $(m-m_c)/m_c$, Fig.~\ref{fig:scaling}B, where $m_c$ increases with increasing $N$ as $m_c\equiv {N}^\kappa$ with $\kappa=0.47\pm 0.02$. This is in a good agreement with the prediction of Eq.~\eqref{eq:capacity} that the memory capacity scales as $m_c\sim {N}^{0.5}$, for the square lattice model with $z=4$ nearest neighbors (see SI Text).

Finally, we have also assessed the trade-off, expressed in Eqn~\eqref{eqn:tradefoff}, between the stability of the multifarious assembly mixture, i.e., its lifetime $t_*$, and the minimal size $N_*$ of a seed needed for retrieval (Fig~\ref{fig:tPlots}).
The minimal seed size $N_*$ increases slowly with increasing $t_*$, and remains a small fraction of the total number of components ($400$, in this case) in a stored structure. The number of stored structures $m$ has only a modest effect on $N_*$, in agreement with Eqn~\eqref{eqn:tradefoff} (see SI Text and SI Fig.~S7).

\subsection{Discussion}
To conclude, we have demonstrated that it is possible to store multiple structures in a solution of components with designed interactions between them. Using $N$ different component species, we can store as many as $\sim (N/M)N^{(z-2)/z}$ different multifarious structures of size $M$ and of average coordination number $z$. In an extended region of parameter values (e.g., temperature, chemical potentials, binding energies), such a ``multifarious assembly mixture" with many stored structures is both stable and responsive; each of the multifarious structures can be selectively grown (retrieved) by modifying chemical potentials or binding energies of only a small fraction of the $N$ component types, or by introducing an appropriate seed. 

\begin{figure*}[h!]
\centering
\includegraphics[width=0.9\linewidth]{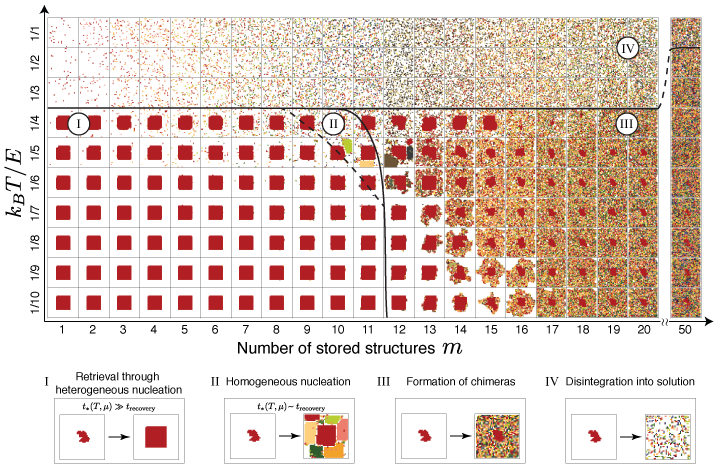}
\caption{Diagram of the different simulation outcomes as a function of the number of stored structures $m$ and temperature $k_BT/E$, starting from a particular supercritical seed (shown in bottom panels). We use different colors to visualize different stored structures, with the seeded structure colored in dark red. Bottom panels distinguish the four regimes identified in the diagram. In regime I the desired structure is retrieved through heterogeneous nucleation since the solution remains stable in the time required for assembly. The solution in this regime is a functional multifarious assembly mixture. Regime II is characterized by homogeneous nucleation of all structures due to reduced stability of the solution (see SI Text Section 3.2). In regime III, formation of structures is dominated by chimeras. Finally, in regime IV, any initial seed is disintegrated into the solution (see also SI text and SI Fig.~S8). These simulations were run for a fixed length of time, $2*10^6$ lattice sweeps, and with fixed chemical potential, $\mu=1.80E$, for all species. The value of $\mu$ mostly influences the extent of regimes I and II. In each plotted snapshot only neighboring tiles that have specific binding between them are plotted, hence tiles without any bonds are omitted. Note that in a system with fixed concentrations, rather than $\mu$, most components would clump to the seed in regime III, while in regime I they would disperse in the solution independently of the structure nucleated from the seed.}
\label{fig:PD}
\end{figure*}

\begin{figure*}[h!]
\centering
\includegraphics[width=0.95\linewidth]{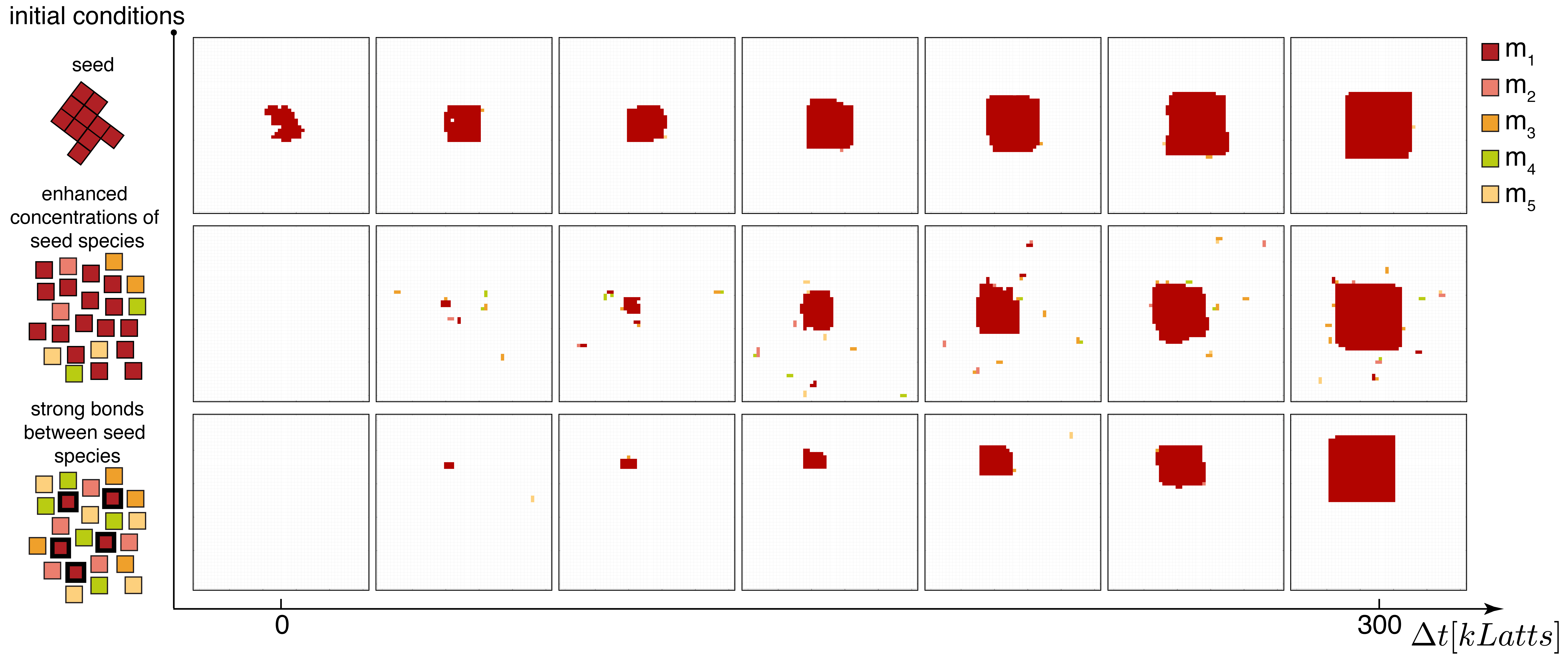}
\caption{Configurations observed during the simulations of retrieval of the desired structure (dark red) in a solution of tiles, whose interactions encode five different structures ($m=5$). Each type of simulation was run at a fixed temperature $k_BT/E=0.15$. The retrieval uses three different triggers: (Top row) a nucleation seed --- i.e., a subcluster of the desired structure, as appears in the first snapshot. We used chemical potential $\mu=1.85E$ for all tile species and observed the retrieval of the desired stored structure progress during the time window between $10^5$ and $3\cdot 10^5$ lattice sweeps; (Middle row) enhanced concentrations of a small number of tile species that can make the seed used in the top row, by using $\mu=1.35E$ for these tile species and $\mu=1.85E$ otherwise. We observed the retrieval of the desired structure in the time between $4\cdot 10^5$ and $7\cdot 10^5$ lattice sweeps; (Bottom row) stronger binding energies $U_{\alpha\beta}=-2E$ between the small number of tile species that can make the seed shown in the top row. We used chemical potential $\mu=1.85E$ for all tile species and observed the retrieval progress in the time between $3\cdot 10^5$ and $6\cdot 10^5$ lattice sweeps. In each of the three rows the snapshots were taken within the time interval $\Delta t=3\cdot 10^5$ lattice sweeps, starting at the time when retrieval begins.}
\label{fig:diffSA}
\end{figure*}

\begin{figure*}[h!]
\centering
\includegraphics[width=0.5\linewidth]{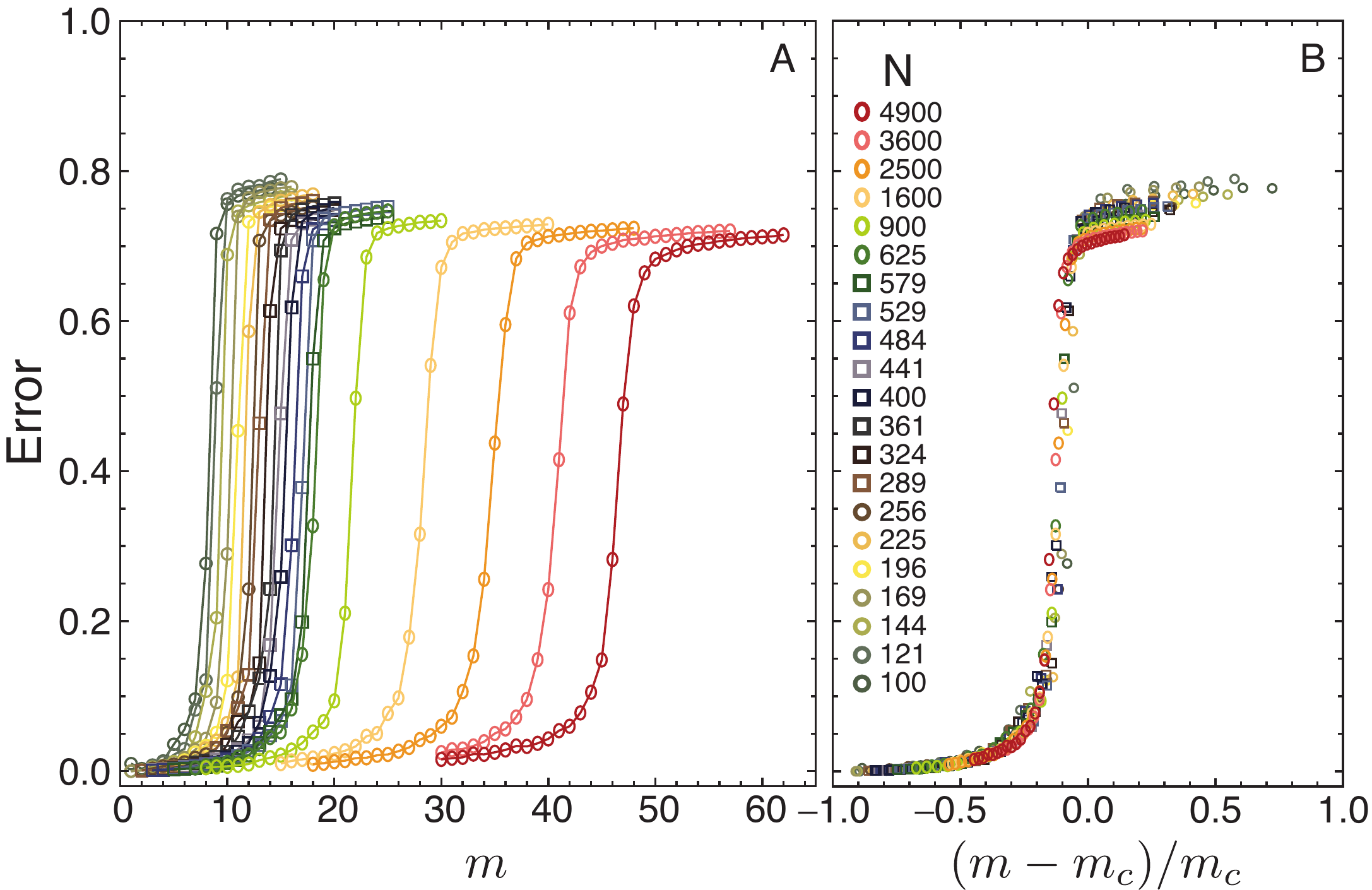}
\caption{Scaling of the storage capacity. A) Measure of difference between the desired structure and the obtained structure (see SI Text and SI Fig.~S4 for the definition and SI Fig.~S1) as a function of the number of stored structures $m$, for different numbers of tile species $N$. Stored structures contain $M=N$ tiles, each of different species. Each point is an ensemble average result of $\sim 100$ different simulation runs. (The curves saturate at $\approx 0.75$, see SI Text for details.) B) Collapse of the curves when the number of stored structures $m$ is rescaled as $(m-m_c)/m_c$, with $m_c\sim {N}^{\kappa}$ and $\kappa=0.47\pm 0.02$. (See SI Text for scaling analysis.) All the simulations were run for values of $N\in[100, 4900]$, with $\mu=1.85E$ and $k_BT/E=0.15$, for a fixed time of $t_{\text{run}}=2*10^6$ lattice sweeps. A diagram of the different simulation outcomes as a function of the number of stored structures and temperature for $\mu=1.85E$ can be found in SI Text.}
\label{fig:scaling}
\end{figure*}

\begin{figure*}[h!]
\centering
\includegraphics[width=0.5\linewidth]{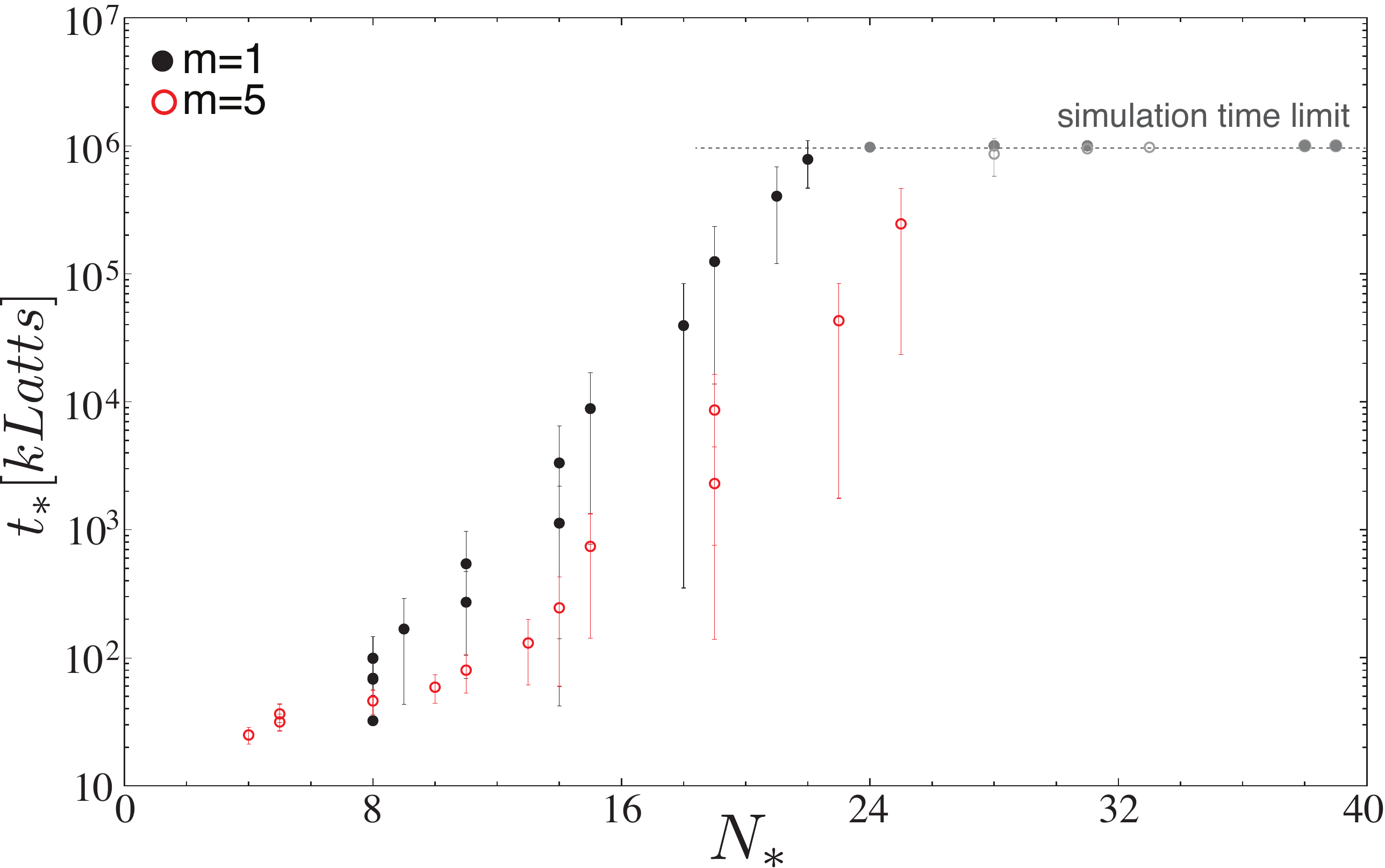}
\caption{Stability of the solution. The characteristic time $t_*$ is plotted here as function of the number of tiles in a critical seed, $N_{*}$, which was varied in simulations by changing the value of $\mu\in[1.50E,1.90E]$. The number of tiles needed to trigger assembly depends weakly (approximately logarithmically) on $t_*$ and remains small compared to the structure size of $M=400$. Note that $N_{*}$ depends very weakly on the number of stored structures $m$. The simulations were run at fixed temperature $k_BT/E=0.15$.}
\label{fig:tPlots}
\end{figure*}

The model that we have explored is very similar to the way associative neural networks, such as Hopfield's classical networks \cite{1982Hopfield} store multiple memories in a distributed way. In these models, a neural network is programmed to have multiple stable states, i.e., memories, using a prescription for neuronal connections that is very similar in spirit to the pooled energy matrix in Eqn~\eqref{eqn:pooledU}. It has been shown \cite{Hertzbook} that if the number of programmed memories is sufficiently small, each memory is indeed a stable state and can be recovered through initial conditions in a robust manner. However, if the number of stored memories exceeds the capacity of the network, recovery is spoiled by the presence of many ``spurious memories'' -- undesired stable states  -- resulting in regimes \cite{Amit:1985tj} similar to those shown in Fig.~\ref{fig:PD}. A distinctive feature of multifarious assembly mixtures, however, is that we require the stability of the unassembled mixture itself for a long time $t_*$, in addition to the stability of the stored structures (see SI Text). 

In our simulated lattice model, different stored structures have identical components rearranged in random permutations. Thus stored structures are assumed to be independent, or ``orthogonal", as in the case of stored memories in the original Hopfield model \cite{1982Hopfield}. An important extension of the present model would be to study stored structures with built-in correlations, such as the presence of shared modules. After all, the controlled assembly of chimeric structures could be useful.
It is also important to stress that designing specific binding interactions between different components based on superposition (Eqn~\eqref{eqn:pooledU}), is not the only way to create functional multifarious assembly mixtures. Although it is arguably the simplest prescription that works for generic structures, other non-linear prescriptions can be tailored for particular structures by exploiting structural motifs (e.g., creating multifarious assembly mixtures with higher capacity or longer lifetimes). Such tailored interactions have been used to store and retrieve a particular set of structures composed of a small number of component species in recent work on DNA programmed assembly \cite{Barish:2009p5463}. In similar vein, the ability of a protein sequence to code for multiple stored internal structures has been studied in the context of protein folding \cite{Fink:2001ki}.

Beyond immediate applications to artificial systems with controllable binding specificity, the present model proposes a new paradigm to understand molecular aggregates in biology. For instance, our calculations show that instead of creating new proteins for every individual structure, it is more efficient if individual proteins are used in a multiplicity of structures, as is the case in many cellular assemblies, ranging from transcription factors \cite{Ihmels,Remenyi} to ribonucleoproteins such as spliceosomes \cite{Wahl2009701}. Our calculations also indicate that such versatility can be quite high, increasing rapidly with the number of different component species in the pool. Nonetheless, different structures can be selectively assembled by reprogramming molecular interactions, e.g. by a simple modulation of the expression levels, or of the specific binding energies via post-translation modifications, of a small number of selected components. This is indeed what seems to happen often in cellular assembly. We hope that the theoretical framework presented here, properly generalized to far-from-equilibrium situations, will form a basis for quantitative studies of functioning, regulation and evolution of biological assembly.


\newpage

\begin{acknowledgments}
We would like to thank our colleagues for discussions and their comments on the manuscript, in particular John Hopfield, David Huse and Olivier Rivoire. Z.Z. acknowledges support from the George F. Carrier Fellowship. M.P.B. acknowledges funding by the National Science Foundation through the Harvard Materials Research Science and Engineering Center (DMR-0820484), the Division of Mathematical Sciences (DMS-0907985), and by Grant RFP-12-04 from the Foundational Questions in Evolutionary Biology Fund. M.P.B. is an investigator of the Simons Foundation.
\end{acknowledgments}



\newpage
\newpage

\section*{Supplementary Information}


\section{Storage Capacity}

\subsection{Promiscuity of Interactions} 
In our general arguments, we assumed that species $\alpha$ and $\beta$ interact specifically if they are bound together in any one of the $m$ stored structures:
\begin{equation}
U^{tot}_{\alpha \beta} = max(\sum_{S=1}^m U^{S}_{\alpha\beta},-E). 
\label{eqn:pooledU}
\end{equation}
This superposition form of $U^{tot}$ implies that each species has multiple specifically binding partners. We define the promiscuity of a binding site on a component of a given species as the number of species that can specifically bind to it. In our model with $N$ total species of components and $m$ stored structures of size $M$ each, a given species will occur in about $m \frac{M}{N}$ stored structures. The component will typically be bound to a different species in each of these structures. Thus the promiscuity of a typical binding site is $\sim m \frac{M}{N}$.

\subsection{Scaling of Capacity} 

When the number of stored structures $m$ is large, the species interactions are highly promiscuous. As a result, a given seed may be able to grow non-uniquely by binding distinct combinations of components and thus form chimeric structures. Here we show that there is a sudden onset of chimeras at some $m = m_c$ which defines the capacity. 

To study this, we can specify arbitrary components along the boundary of a seed (black tiles in Fig.~S\ref{fig:SIFig_boundary_conditions}A) and ask if there are multiple choices of species that can fill in the positions $Z_1, Z_2, \ldots$ shown in Fig.~S\ref{fig:SIFig_boundary_conditions}A, such that $Z_i$ form specific bonds with each other and with all the seed components. For example, the number of species that can form a specific bond with the component $13$ from the right (i.e., number of choices for site $Y$ in Fig.~S\ref{fig:SIFig_boundary_conditions}B) is given by the promiscuity $\sim m \frac{M}{N}$ of $13$'s right-side binding site. Similarly, we have another set of $\sim m\frac{M}{N}$ choices of species for site $X$ that bind specifically to the component $14$ from below. The choices for site $Z_1$, i.e. species that can bind specifically to both $13$ and $14$, is given by the intersection of these two sets.

If the $m$ stored structures are randomly and independently constituted from the $N$ species, we can assume that these sets of choices for $X$ and $Y$ are two random uncorrelated sets of size $m \frac{M}{N}$ contained in the set of all $N$ species. We  then estimate the probability of having at least one element in the intersection of the two sets to be,
\begin{equation}
P_{\cap} = 1 - \left(1- \frac{m M}{N^2}\right)^{\frac{m M}{N}}.
\end{equation}

To estimate the probability that at least one extended chimeric structure of length $L$ (like that shown in red in Fig.~S\ref{fig:SIFig_boundary_conditions}A) exists, we assume a choice of species for site $Z_1$ that is stably bound and compute the probability of stably-bound chimeric choices for $Z_2$, fix a choice for $Z_2$ and compute choices for $Z_3$  and so on. Such an estimate of the probability of existence of an extended chimeric structure of length $L$ is given by:
\begin{equation}
P_{L} \sim P_{\cap}^L = \left(1 - \left(1- \frac{m M}{N^2}\right)^{\frac{m M}{N}}\right)^L.
\end{equation}
Taylor expanding  for small $\frac{m M}{N}$ gives
\begin{equation}
P_{L} \sim \left(\frac{m^2 M^2}{N^3}\right)^L.
\end{equation}
For large $L$, this probability sharply grows at $m \approx m_c$, where 
\begin{equation}
m_c \sim \frac{N}{M} \sqrt{N}.
\end{equation}

A similar argument applies for a general lattice structure with coordination number $z$. In this more general case, a component occupying site $Z_i$ must form specific bonds with $z/2$ boundary components. Hence the choice of species for each $Z_i$ is given by the intersection of $z/2$ sets of size $\frac{m M}{N}$ each, so that the Taylor expansion for $P_{\cap}$ is modified to $P_{\cap} = N \left(\frac{m M}{N^2}\right)^{\frac{z}{2}}$. The formula for capacity with general coordination number $z$ is therefore:
\begin{equation}
m_c \sim \frac{N}{M} N^{(z-2)/z}.
\end{equation}

\subsubsection*{Explicit counting of chimeras using transfer matrices}
 
We can also explicitly count the number of stably-bound chimeric structures $\nu_L(m,N,M)$  of length $L$, i.e., the number of choices for the set $\{ Z_1, \ldots , Z_L \}$ shown in Fig.~S\ref{fig:SIFig_boundary_conditions}. Thus, $P_L$ estimated above is the probability that $\nu_L \geq 1$.  

We can numerically compute $\nu_L$ for a square lattice through explicit enumeration using a transfer-matrix-like method \cite{Baxter}.  

Let $w_a$ be the set of species that can occupy site $Z_a$ in Fig.~S\ref{fig:SIFig_boundary_conditions}A by forming a specific bond with the boundary component from the right. (For instance,  $w_{2}$ is the set of species that specifically bind to the component species $10$ from the right.) We form a ``transfer matrix'' $T_{a}$ between sites $a$ and $a+1$ by restricting the matrix $e^{-\beta U^{tot}_{\alpha \beta}}$ to rows $\alpha$ which are species found in $w_a$ and to columns $\beta$ which are species found in $w_{a+1}$. Hence $T_{a}$ is a $|w_a| \times |w_{a+1}|$ matrix of Boltzmann factors for binding between species $w_a$ and $w_{a+1}$ that can stably occupy sites $a$ and $a+1$. (Note that we use the interaction matrix $U^{tot}$ between top and bottom faces, i.e., vertical direction in Fig.~S\ref{fig:SIFig_boundary_conditions}A, in this construction of $T_a$.) Then, the sum $\zeta_L$ of all entries of the matrix product,
\begin{equation}
\zeta_L = \sum \sum T_1 T_2 \ldots T_L
\end{equation} gives the partition function summed over all chimeric structures made of components ${Z_1,\ldots,Z_L}$ which, by construction, form specific bonds with the (black) boundary components to the left in Fig.~S\ref{fig:SIFig_boundary_conditions}A. The most stable of these chimeric structures will also contain specific (vertical) bonds between every pair $Z_a, Z_{a+1}$, i.e., $L-1$ specific vertical bonds of energy $-E$ each. Hence, if we multiply the partition function $\zeta_L$ by $e^{+\beta (L-1) E}$ and take the limit $\beta \to \infty$, only terms corresponding to chimeric structures with $L$ specific vertical bonds will survive. In fact, in the large $\beta$ limit, $e^{+\beta (L-1) E} \zeta_L$ precisely gives us the number $\nu_L$ of such stably-bound chimeric structures.

Using this method, we can compute $\nu_L$ explicitly for given boundary conditions. Averaging this count over $200$ realizations of $m$ random structures of size $M=N$, we obtained $\nu_L(m,N)$ shown in Fig.~S\ref{fig:SIFig_boundary_conditions}C. Here we chose  $L \sim \sqrt{N}/2$ to only count chimeric structures of length comparable to the side length of the $\sqrt{N} \times \sqrt{N}$ structure itself. (However, any $L$ that grows with $N$ gives similar results, as supported by the probabilistic argument for $P_L$ above.) By varying $N$ between $36$ and $10,000$, the inset of Fig.~S\ref{fig:SIFig_boundary_conditions}C shows that $\nu_L$ rises rapidly at $m=m_c \sim \sqrt{N}$. 

Both our probabilistic arguments, using $P_L$ and explicitly counting $\nu_L$ of chimeric structures, agree on the scaling of capacity $m_c \sim \sqrt{N}$. These results also agree with the scaling extracted from Monte-Carlo simulations of the lattice model, presented in Fig.~4 and in Section~\ref{sec:scalingsim} below.

\section{Minimal seed size and lifetime of the multifarious assembly mixture}

We require two distinct behaviors of the multifarious assembly mixture:

\begin{itemize}
\item \textbf{Responsive:} The multifarious assembly mixture must produce structures in response to an externally introduced seed (or equivalent perturbation) of small size,
\item \textbf{Stable:} The multifarious assembly mixture must be stable in the absence of external signals and must not produce any structures spontaneously. 
\end{itemize}
Nucleation theory, adapted to multifarious structures, dictates whether such competing requirements can be implemented.

\subsection*{Minimal seed size needed $r_*$.}

We begin with the question of how large a seed is needed to recover a stored structure. The change in free energy, with respect to the multifarious assembly mixture, due to the presence of an $r \times r$ square seed taken from one of the $m$ stored structures is:
\begin{equation}
F(r) = 2 E r (r-1) - \mu r^2,
\label{eq:fr}
\end{equation}
since such a structure has $r^2$ components and $2 r (r-1)$ strong bonds. Here $-E < 0 $ is the energy of specific bonds and $\mu >0$ is the chemical potential of each species. (Note that we neglect the change of entropy in Eqn.~\eqref{eq:fr}.)

The general shape of $F(r)$ is shown in Fig.~S\ref{fig:SIFig_NucleationBarriers}; as in conventional nucleation theory, $F(r)$ has a maximum for some critical size $r_*$. A sub-critical seed of size $r < r_*$ will dissolve back into its components while supercritical seeds (i.e., size $r > r_*$) will grow in size and into the full stored structure.

Hence the minimal size of the seed we must introduce to recover structures is simply given by the size $r_*$ of the critical seed. We can calculate $r_*$ by setting $\partial_{r} F = 0$. We find
\begin{equation}
r_* = \frac{E}{2E-  \mu},
\label{eq:rstar}
\end{equation}
with $F_* \equiv F(r_*) = (2 E - \mu) r_*^{2}$. Note that this relationship is independent of the number of stored structures $m$; the minimal seed size is determined by a local condition $\partial_r F = 0$ on the free energy landscape and is not affected by the presence of other minima.

\subsection*{Lifetime of the multifarious assembly mixture $t_*$.}
The excess free energy potential $F(r)$ shown in Fig.~S\ref{fig:SIFig_NucleationBarriers} implies that the multifarious assembly mixture is intrinsically unstable ---  even without the external introduction of any seed, a critical seed of size $r_*$ could spontaneously emerge on some timescale $t_*$, leading to the nucleation of random stored structures. Hence the timescale of such spontaneous nucleation $t_*$ sets the useful lifetime of the multifarious assembly mixture.

In conventional nucleation theory, the timescale for spontaneous nucleation $t_*$ is given by Arrhenius's formula for barrier crossing $t_* \sim e^{-\frac{F_*}{k_B T}}$. However, with $m$ multifarious structures, we need to modify this formula to account for multiple inequivalent seeds that can spontaneously nucleate distinct stored structures. If there are $q(m,N,M)$ inequivalent barriers that can be crossed, the timescale for spontaneously crossing any one of the barriers and assembling a stable stored structure is given by,
\begin{equation}
\log \frac{t_*}{\tau} = \frac{F_*}{k_B T} - \log q(m,N,M),
\label{eqn:tstarFstar}
\end{equation}
where $\tau$ is a timescale associated with microscopic processes. In conventional nucleation theory, there is only one (or $O(1)$) stable phase, while seeds can vary only in shape and not in composition. 
In contrast, our multifarious assembly mixture can form at least $m$ stable structures (i.e., the $m$ stored structures) in addition to any stable chimeric structures that might exist. Further, seeds from different parts of these multifarious structures are inequivalent in composition. 

For small $m \ll m_c$, we can ignore chimeric structures and estimate
\begin{equation}
q \sim m M,
\label{eq:q}
\end{equation}
since the $m$ stored structures can each be nucleated with $\sim M$ inequivalent seeds. 
We do not pursue the detailed form of this correction any further here; the correction to Eqn.~ \ref{eqn:tstarFstar} is logarithmic and numerical simulations discussed below confirm that the $\log q$ correction is modest. 
 
Using the expression for $F_*$, we find,
\begin{equation}
\log \frac{t_*}{\tau} = (2 E - \mu) r_*^{2} - \log q(m,N,M).
\label{eq:tstar}
\end{equation}
Note that we have an unavoidable trade-off: increasing the lifetime of the multifarious assembly mixture $t_*$ would necessarily increase the critical seed size $r_*$. Thus a more stable multifarious assembly mixture requires a larger seed for recovering stored structures.

We can also rewrite the lifetime $t_*$ in terms of the parameters $\mu$ and $E$ as
\begin{equation}
\log \frac{t_*}{\tau} = \frac{E^2}{k_B T (2 E - \mu)} - \log q(m,N,M),
\label{eq:tstar2}
\end{equation}
and we tested this relationship in our Monte Carlo simulations.

\section{Monte Carlo simulations}

We carry out Monte Carlo simulations of this system, with $N$ different species of square tiles on a square grid, see Fig.~\ref{fig:simsetup}. Each stored structure is a $\sqrt{M}\times\sqrt{M}$ sized square composed as a random permutation of $M\equiv N$ tiles, with each tile being of different species. Since we work in the grand canonical ensemble, the total system was larger, with a side length $\sqrt{N_0}=2\sqrt{M}$, when $\sqrt{M}$ was even, and $\sqrt{N_0}=2\sqrt{M}+1$ otherwise.

The energy of a state of the system is specified by  $U^S_{\delta}$, where $S\in\{1,\ldots,m\}$ labels the stored structures, while $\delta=1\ldots z$ labels the bond directionality of the nearest neighbor tile positions, with $z=4$ being the coordination number (or valance). For our square grid, $\delta=1\ldots 4$ labels up, right, down and left neighbor positions, respectively. For a given structure $S$ we consider all nearest neighbor pairs of tiles. When a pair of tiles of species $\alpha,\beta$ is found, $\delta$ is determined as the directionality of the distance vector $\vec{r}_{\alpha\beta}$, and $U^S_{\alpha\beta,\delta}$ is set to $-E$. The elements of the total interaction matrix for the system are:

\begin{equation}
U^{tot}_{\alpha \beta,\delta} = max(\sum_{S=1}^m U^S_{\alpha\beta,\delta},-E). 
\label{eq:1}
\end{equation}
The maximum function caps the matrix elements, since we consider a model with all bonds of the same strength.

Finally, we define the energy of states. The square grid has $N_0$ sites, each being either empty or occupied by a tile which can be one of $N$ different species. An empty site can be simply treated as a tile of zeroth species, which has no binding energy, so $U_{\alpha\beta,\delta}^{tot}\equiv0$ when any of $\alpha,\beta$ is zero, and with chemical potential $\mu_0=0$. A state of the system is then described by a vector array $\vec{\sigma}_i=\{\sigma^\alpha_i\}$, where $\sigma^\alpha_i=1$ if $\alpha\in\{0,1,\ldots, N\}$ is the tile species at grid site $i=1\ldots N_0$, and $\sigma^\alpha_i=0$ otherwise, with $\sigma^0_i=1$ labeling site $i$ void of tile.
The energy of such a state is:
\begin{align}
  U&=\sum_{\langle i,j\rangle}\vec{\sigma}_i\cdot U^{tot}_{\delta(i,j)}\cdot\vec{\sigma}_j + \sum_{i}^{N} \vec{\mu}\cdot \vec{\sigma}_i\\\notag
  &=\sum_{\langle i,j\rangle}\sum_{\alpha,\beta}U^{tot}_{\alpha\beta,\delta(i,j)}\sigma^\alpha_i\sigma^\beta_j + \sum_{i}^{N}\sum_\alpha^{N} \mu_\alpha\sigma^\alpha_i,
  \label{eq:MCenergy}
\end{align}
where $\delta(i,j)$ is the directionality defined by nearest neighbor pair $\langle i,j\rangle$, and $\vec{\mu}=\{\mu_\alpha\}$ the vector of chemical potentials for species $\alpha$.

The Monte Carlo algorithm we used in the simulations chooses a random grid position and changes its species with a probability $e^{-\Delta U/k_B T}$, where $\Delta U$ is the total energy cost of the change calculated using Eq.~(15). Typical simulations were run for $10^6Latts$, where $1Latts$ is one lattice sweep, i.e., $N_0$ Monte Carlo moves.

\subsection{Scaling of capacity in simulations}
\label{sec:scalingsim}

Fig.~4A shows the error in an assembled structure, observed at the end of simulation, as a function of the number of stored structures $m$, for different structure sizes $M$ (and correspondingly different system sizes $N$ and numbers of tile species $M=N$). Each simulation starts with one selected complete structure of a given size $M$ (in shape of a square) and runs for a fixed amount of time $t_{\text{run}}$. We define the error using a three-step procedure, see Fig.~S\ref{fig:errordef}: first, in the final state of the simulation we identify the largest contiguous area of tiles that are bonded, i.e., the largest connected structure, which we call the "final structure"; next, the union of the area of the final structure with the area of the initial structure (which is a square) gives a total area $A$, and the number of tiles that match between the initial and final structures inside $A$ is divided by the total number of tiles in $A$ to give the overlap of structures. By definition, this overlap is between zero and one. However, since the initial structure in the simulations never dissolves in the considered regimes (I and III), the overlap is at least $M/N$ (for the system sizes we studied $M/N\approx 0.25$). Finally, the error is defined as one minus the overlap. This definition of error, which uses $A$, consistently takes into account the multifarious assembly mistakes that occur by changing or loosing tiles in the initial structure, as well as by attaching tiles to the boundary of the initial structure.

For any given structure size $M$, the error in the simulations sharply rises from zero when a certain number of stored structures was reached, and then quickly saturates at $1-M/N\approx 0.75$. The capacity $m_c$ is the number of structures that can be stored without having a large error, and we extract its value using a finite size scaling analysis. Our Ansatz for the scaling function is $\text{Error}=f(\frac{m}{m_c})$, with $m_c\equiv N^\kappa$, where $m$ is the number of stored structures and $\kappa$ is the only fit parameter in function $f$. We find it robust for analysis to consider only the datapoints for which $\text{Error}\leq 0.5$, which is consistent with our interest in the sharp rise of error from zero. We rescale the datapoints in each error curve  using $m\rightarrow\frac{m}{N^\kappa}\equiv x$, with $\kappa$ fixed, and we fit the complete set of datapoints  using a polynomial of fifth degree which represents the unknown function $f(x)$. We quantify the quality of the fit for the given $\kappa$ with $\chi^2=R/(r+p)$, where $R$ is total of squared fit residuals, $r$ is total of squared mean prediction errors, and $p$ is total of squared datapoint errors \cite{PhysRevB.70.014418}. We minimize $\chi^2$ with respect to $\kappa$. The $\chi^2$ varied smoothly with $\kappa$ and its minimum is easily found, giving the optimal value of $\kappa$. Fig.~4b shows a very good collapse of error curves for different number of species $N$ when the optimal $\kappa$ is used (the parts of the curves with $\text{Error}>0.5$ are also plotted), which validates the scaling Ansatz.

Finally, we describe the procedure for estimating the error of the calculated optimal value of $\kappa$. We use a simple bootstrap method \cite{barkemaBook}. Consider the error curve for some system size, and let there be $n$ datapoints in this curve. We form a new dataset by randomly choosing a datapoint from the curve $n$ times. The new curve therefore has the same number of points as the original, but some of the original points might be missing and some might occur multiple times. This sampling is applied to every error curve. Using the new datasets, the new optimal value of $\kappa$ is calculated. We repeat the entire procedure  a hundred times (always starting from the original datasets), giving a distribution of optimal $\kappa$ values. The mean of the distribution is the $\kappa$ quoted in the main text, while the standard deviation is its error.

We check that using higher order polynomials in the fit procedure does not significantly improve the quality of fits. In addition, instead of taking the error cutoff equal to $0.5$, we also vary it between $0.2$ and $0.7$, but the variation in obtained optimal $\kappa$ was of the same order of magnitude as the error obtained by the bootstrapping procedure.

\subsection{Minimal seed radius $r_*$ and lifetime $t_*$ of the multifarious assembly mixture in simulations}\label{sec:rtstarsim}
To test the relationship between $r_*$, $\mu$ and $m$ (Eqns.~\eqref{eq:rstar} and \eqref{eq:tstar2}), we ran a series of simulations starting with a square-shaped seed of different sizes and various values of $\mu$, and measure the probability that the seed dissolves. We consider $m=1$ and $m=5$ stored structures at fixed temperature $T=0.15$, well inside the retrieval regime.

Fig.~S\ref{fig:diss} shows the dissolving probability as a function of the number of tiles in a seed, $N_{seed}$, each given seed approximating a disc in shape. Each probability averages $100$ simulation runs. From this data we extract the critical radius of the seed, $r_*$, at which the probability to dissolve drops to zero. We show the extracted $r_*$ as a function of $\mu$  in Fig.~S\ref{fig:rstarMu}. Note that in this section we redefine the quantity $r_*$ as dimensionless and equals the critical seed radius measured in the units of tile length.

\subsubsection*{Lifetime $t_*$ in simulations}
We investigate different time scales in the Monte Carlo simulations of the lattice model. Fig.~S\ref{fig:tstarMum}A plots both the spontaneous nucleation time $t_*$ and the recovery time $t_{\text{rec}}$ measured as functions of the chemical potential $\mu$. We define the spontaneous nucleation time $t_*$ as the time when any structure, being spontaneously assembled in the homogeneous solution, reached an area of $M/4=100$ tiles, which is above the critical nucleation size for the largest value of the chemical potential we considered (see Fig.~S\ref{fig:diss}). Recovery time $t_{\text{rec}}$ identifies the moment in the simulation when we first observed the seeded structure completely assemble (seed had $79$ tiles, which is supercritical for all chemical potentials considered). In some simulations, at the moment of completed assembly, there are a few erroneous tiles attached to the structure, which we neglect here.

As $\mu/E\to 2$, the $t_*$ diverges, while the recovery time is much smaller and essentially independent of $\mu$. This demonstrates that there is a parameter range of the model where structure retrieval occurs much more quickly than spontaneous nucleation from the solution, Fig.~S\ref{fig:tstarMum}A. In Fig.~S\ref{fig:tstarMum}B shows how the nucleation time $t_*$ depends on the number of stored structures. The obtained result is consistent with Eqns.~\ref{eq:q} and \ref{eq:tstar2} of the main text.

\subsection{Transition from chimeric regime to homogeneous solution regime in simulations}

Here we present details of the finite temperature transition between regimes III and IV presented in Fig.~2. As presented in Fig.~S\ref{fig:transitionT}, we pick the simulations with $m=17,30$ and $50$ stored structures as examples, and look at the size of largest structure as function of temperature. At the end of each simulation we identify the ``biggest structure'' as the largest contiguous area of tiles that are bonded specific interactions, i.e., the largest connected structure. At each temperature we average the size (number of tiles) of the biggest structure  over $10$ independent runs. Fig.~S\ref{fig:transitionT} reveals that the biggest structure covered most of the entire system at low temperatures (regime III), but its size sharply dropped at a certain temperature. At temperatures above this transition (regime IV), the system is mostly filled with tiles, however, there are hardly any specific interactions between them, i.e., the state was a solution of fluctuating components. As expected, the transition temperature increases slowly with increasing number of stored structures to $-E/k_BT=-1$, with $-E$ the binding energy.

\subsection{Regimes observed in Monte Carlo simulations}

In Fig.~S\ref{fig:regimesmu}A-B we show regimes observed in Monte Carlo simulations as a function of the number of stored structures $m$ and temperature $k_BT/E$, starting from a particular supercritical seed (shown in bottom panels). As in Fig.~2, we use different colors to visualize different stored structures, with the seeded structure colored in dark red. Bottom panels distinguish the four regimes identified in the diagram. In regime I the desired structure is retrieved through heterogeneous nucleation since the solution remains stable in the time required for assembly. The solution in this regime is a functional multifarious assembly mixture. Regime II is characterized by homogeneous nucleation of all structures due to reduced stability of the solution. In regime III, formation of structures is dominated by chimeras. Finally, in regime IV, any initial seed is disintegrated into the solution.  

These figures differ from Fig.~2 in the value of the chemical potential $\mu$, which is the same for all the species. In Fig.~S\ref{fig:regimesmu}A the lifetime $t_*$ of the multifarious assembly mixture is suppressed due to the lower value of $\mu=1.70E$. Hence, regime II suppresses the retrieval regime I compared to the result in Fig.~2, where $\mu=1.80E$.

In Fig.~S\ref{fig:regimesmu}B the chemical potential is higher $\mu=1.85E$. Consequently the characteristic lifetime of the multifarious assembly mixture is $t_*\gg t_{recovery}$ (see Fig.~S\ref{fig:tstarMum}), resulting in complete suppression of regime II.

\section{Connections to neural networks and thermodynamic limits}
\setcounter{figure}{0} \renewcommand{\thefigure}{S.\arabic{figure}}

Our model of multifarious assembly mixtures is closely related to models of associative memory \cite{1982Hopfield}. In these models \cite{HertzBook}, multiple ``memories'' are stored as stable states of a neural network by choosing the connections between $N$ neurons appropriately. Just as with multifarious assembly mixtures, neural networks have a finite capacity; with $N$ neurons, a limited number of memories $m < m_c(N)$ can be reliably stored and retrieved. If the capacity $m_c(N)$ is exceeded, many spurious memories -- undesired stable states -- appear and interfere with retrieval. 

Neural networks have also been studied in the thermodynamic limit of a large number of neurons $N$ \cite{Amit:1985tj,Amit:1985um}. In this limit \cite{Amit:1985tj}, energy barriers separating the memories grow  large and each memory becomes a stable thermodynamic phase of the system, provided the number of memories $m = \alpha N$ is less than a critical value $\alpha_c N$. However, there is a phase transition at $\alpha \sim \alpha_c$ to a spin-glass phase with an exponential number of other spurious but stable states.  The properties of such phases have been worked out for different neural networks with different models of interactions between neurons, resulting in phase diagrams that resemble Fig.~2 of our paper. In fact, a recently studied \cite{Hopfield:2010tk,Monasson:2013ub} $2d$-lattice neural network, albeit with long-ranged interactions, is closely related to the large coordination number $z$ limit of our model. Hence, it is natural to ask about the thermodynamic properties of multifarious assembly mixtures. For example, we can take the size of programmed structures $M \to \infty$ to be large, with the ratio $N/M$ held finite. Such a limit might allow for a growing number of stored structures $m = \alpha (N/M) N^{(z-2)/z}$ to become stable phases of the system, provided $\alpha $ is less than a critical $\alpha_c$. 

However, a crucial intrinsically-kinetic feature of multifarious assembly mixtures, not found in neural networks, is the stability of the unassembled mixture itself. As we showed in the paper, the unassembled mixture has a finite lifetime $t_*$ after which random stored structures are spontaneously nucleated. This lifetime $t_*$ is set by the ratio $\mu/E$ of chemical potential to bond energy and hence, in principle, is independent of structure size $M$ and can stay finite in the thermodynamic limit.

As a result, for multifarious assembly mixtures, the question of practical interest is a finite-time kinetic question --- can one recover specific structures using seeds in a time shorter than $t_*$? To focus on this question, we define capacity and the transition to chimeric phase through a practical finite-time notion of error --- we introduce an initial seed and measured error in recovery after a finite simulation time chosen to be much smaller than the lifetime of the mixture.
 
A thermodynamic analysis of our model would require a modified set of quantities and parameter limits. For example, the unassembled mixture can itself be stabilized as a phase only if its lifetime $t_*$ diverges in the thermodynamic limit, which requires tuning the chemical potential $\mu(M)/E \to 2$ with growing structure size $M \to \infty$. In such a limit, our finite-time error can be replaced by order parameters analogous to those used in \cite{Amit:1985tj,Monasson:2013ub} to study the transition to the chimeric regime. We leave a detailed study of the thermodynamic limit of multifarious assembly mixtures to future work.


\newpage

\begin{figure*}[h!]
\centering
\includegraphics[width=0.65\textwidth]{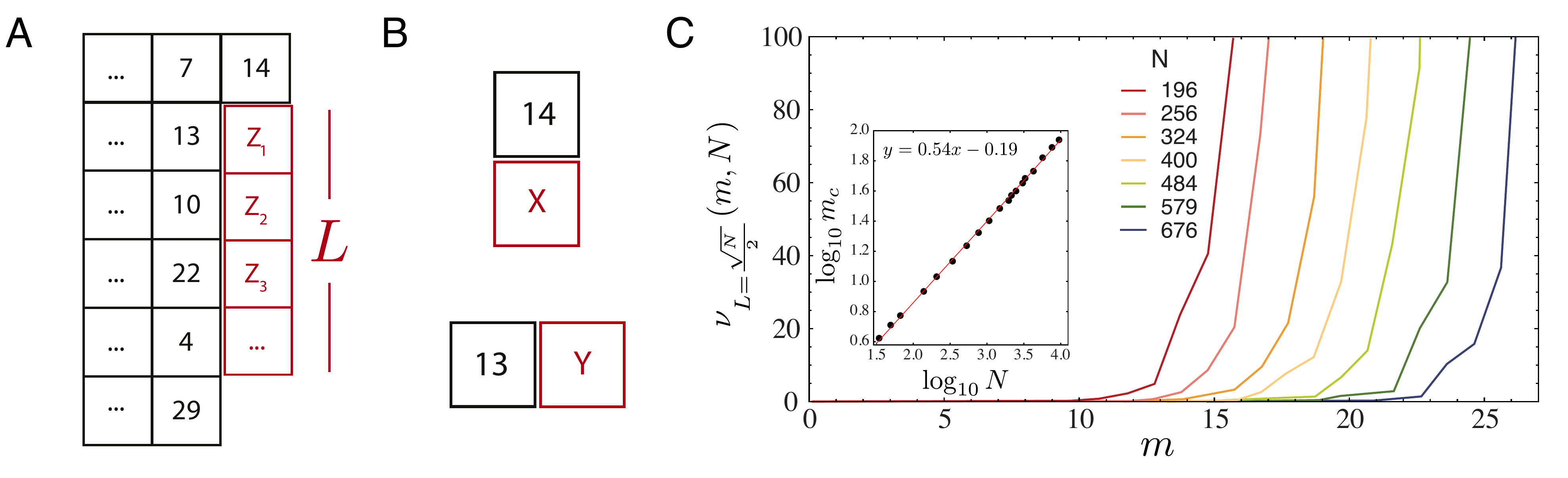}
\caption{A) If a growing seed (black) can bind distinct sets of species through specific bonds, chimeric structures will form. To determine when this happens, we ask how many different choices $\nu$ of species can fill in the red sites $Z_a$ such that they bind specifically to each other (vertically) and to an arbitrary seed boundary (numbered black tiles to the left). Such choices correspond to stable chimeric structures that interfere with recovery. B) The choice of species for $Z_1$ is the intersection of the set of species $X$ that bind specifically to $14$ and species $Y$ that bind specifically to $13$. C) We used a transfer-matrix-like technique to explicitly count the number $\nu_L(m,N)$ of ways of filling in sites $Z_1, \ldots Z_L$ in A) with species such that all bonds (vertical and to the left with the black seed boundary) are specific. Thus $\nu_L$ gives the number of stable chimeric structures that can grow on the seed boundary.  We find that $\nu_L$ rises sharply with number of memories $m$ at $m_c(N)\sim {N}^{\kappa}$, with $\kappa=0.54\pm 0.01$ consistent with the prediction $\kappa=1/2$ (inset). We chose the length $L \sim \sqrt{N}/2$ comparable to the linear size of the $\sqrt{N} \times \sqrt{N}$ shaped stored structures, however the capacity scaling $m_c\sim \sqrt{N}$ is not influenced by this choice but rather by the coordination number of the lattice. \label{fig:SIFig_boundary_conditions}}
\end{figure*}

\begin{figure*}[h!]
\centering
\includegraphics[width=0.5\linewidth]{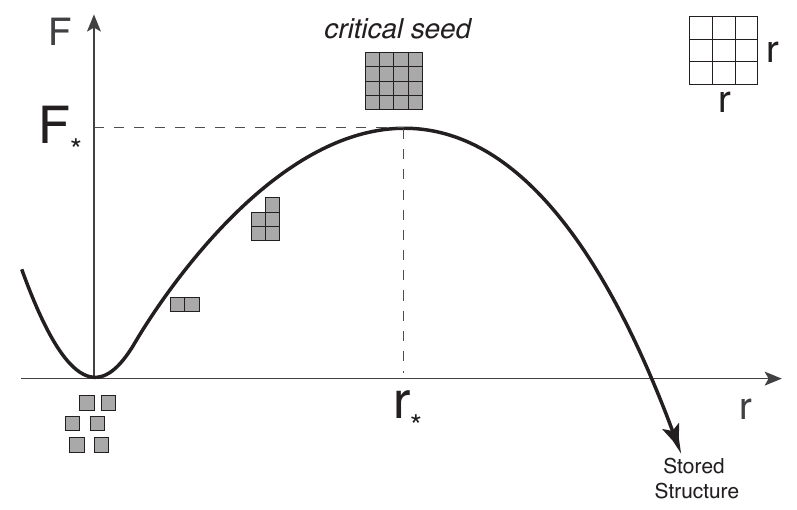}
\caption{The change in free energy $F(r)$ of a $r \times r$ square seed taken from a stored structure. Only seeds with $r > r_*$ will grow into a stored structure. The multifarious assembly mixture of unbound components can also spontaneously nucleate a critical seed on a timescale $t_*$, leading to the assembly of random stored structures without any external cue. Thus $t_*$ sets the useful lifetime of the multifarious assembly mixture. A multifarious assembly mixture with a longer lifetime $t_*$ (i.e., higher stability) necessarily requires larger seeds $r_*$ for recovering structures (i.e., lower responsiveness).}
\label{fig:SIFig_NucleationBarriers}
\end{figure*}

\begin{figure*}[h!]
\centering
\includegraphics[width=0.55\linewidth]{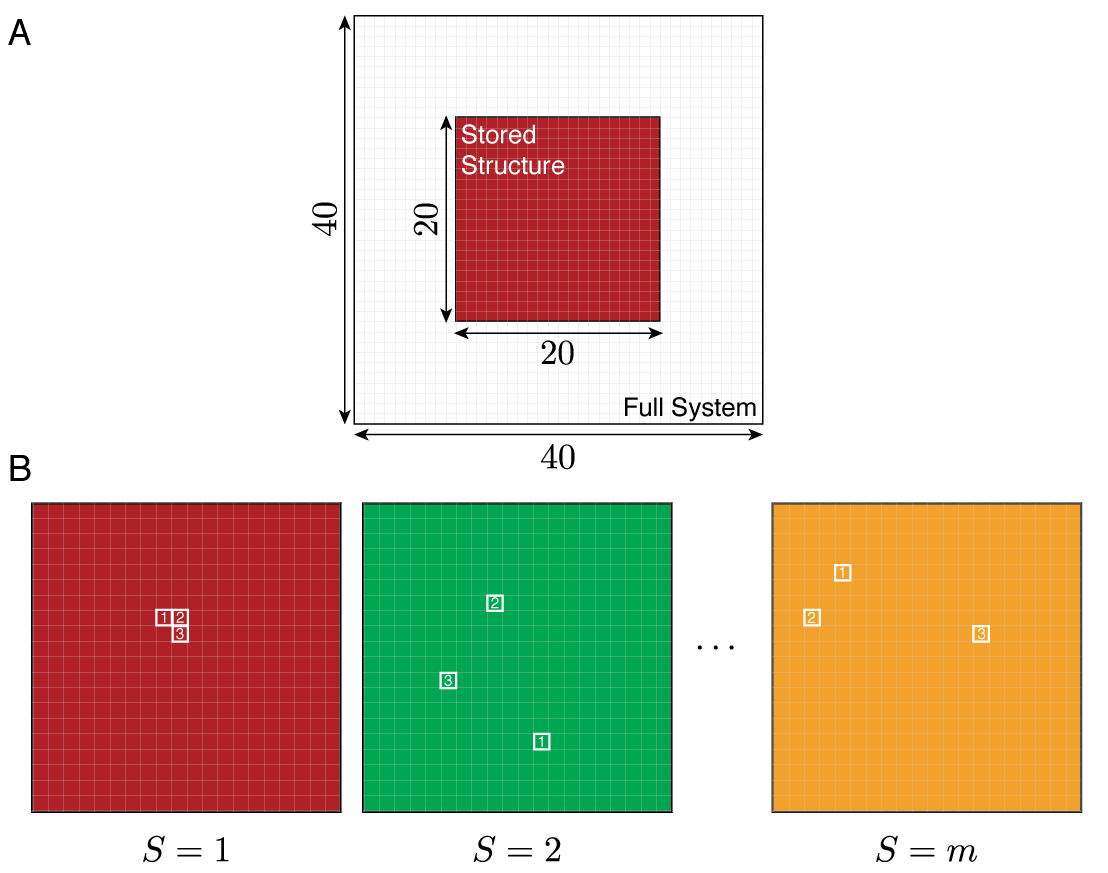}
\caption{\label{fig:simsetup} Monte Carlo simulations on a square grid. A) All stored structures consist of $M_S=400$ tiles forming a square (dark red). The entire system has $N_0=1600$ grid sites which can be occupied by tiles. B) There are $m$ different stored structures indexed by $S$. We take $N_S=400$ tiles, each being of different species (there are $N=400$ species available), and randomly permute them inside the square of size $M_S=400$, to obtain a stored structure. In other words, each tile species occurs exactly once in each structure, with examples of tile species $1,2,3$ depicted, making the species multiplicities $n^S_\alpha=1$, i.e., $N=N_S=M_S=M$.}
\end{figure*}

\begin{figure*}[h!]
\centering
\includegraphics[width=0.25\linewidth]{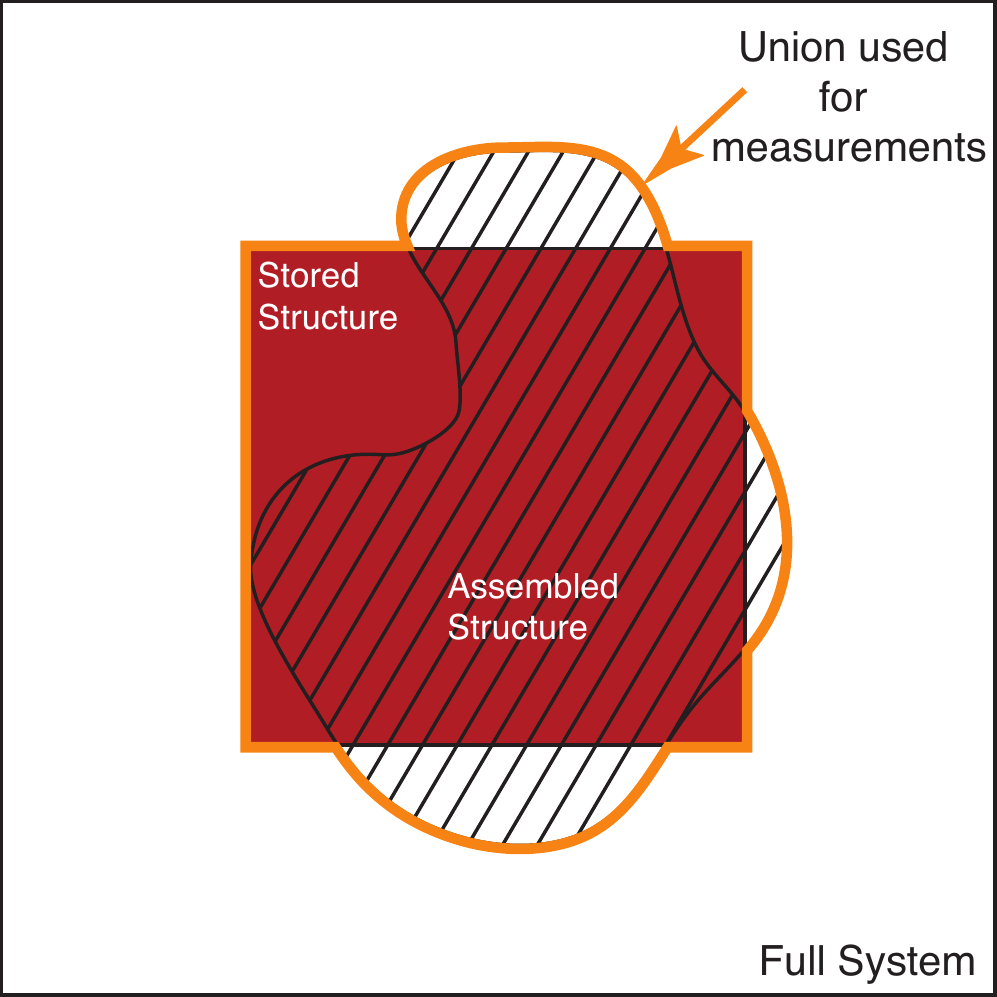}
\caption{\label{fig:errordef} Sketch of how error in assembled structure is defined. The initial structure was a square of size $M$ (dark red). The "final structure" was identified in the final state of the simulation, as the largest contiguous area of tiles that are bonded, i.e., the largest connected structure (shaded). The area $A$ is the union of initial and final structures (bounded by orange line). The number of tiles that match between the initial and final structures inside $A$ is divided by the total area of $A$ to give the overlap of structures. The error is defined as one minus the overlap. This definition of error takes into account all assembly mistakes, such as changed and lost tiles in the initial structure and tiles attached to the initial structure. In considered regimes (I and III), the errors tend to spread the final structure across the entire system of total area $N$ (giving overlap $M/N\approx 0.25$ or more), but rarely shrink it compared to initial structure (which could give overlaps from $1$ down to $0$). The measured error therefore saturated at $\approx 1-0.25=0.75$.}
\end{figure*}

\begin{figure*}[h!]
  \centering
\includegraphics[width=0.95\linewidth]{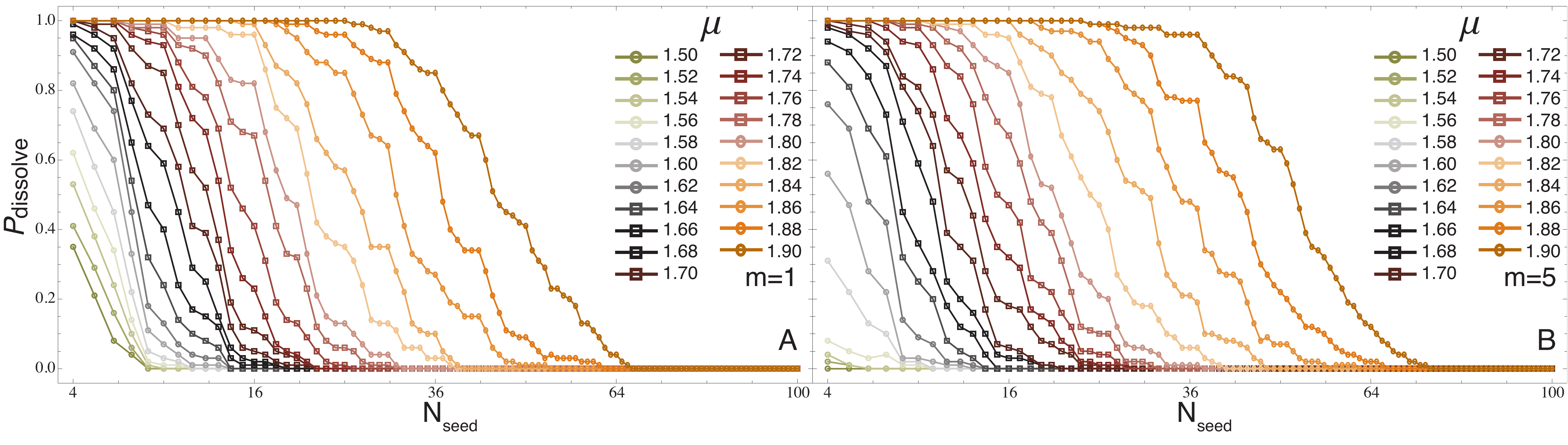}
\caption{\label{fig:diss} Probability to dissolve a seed of size $N_{seed}$ for different values of chemical potential in the range $\mu\in[1.5,1.9]$, and number of stored structures being A) $m=1$, and B) $m=5$. Each point is calculated as a probability based on $100$ independent simulation runs.}
\label{fig:diss}
\end{figure*}

\begin{figure*}[h!]
  \centering
\includegraphics[width=0.5\linewidth]{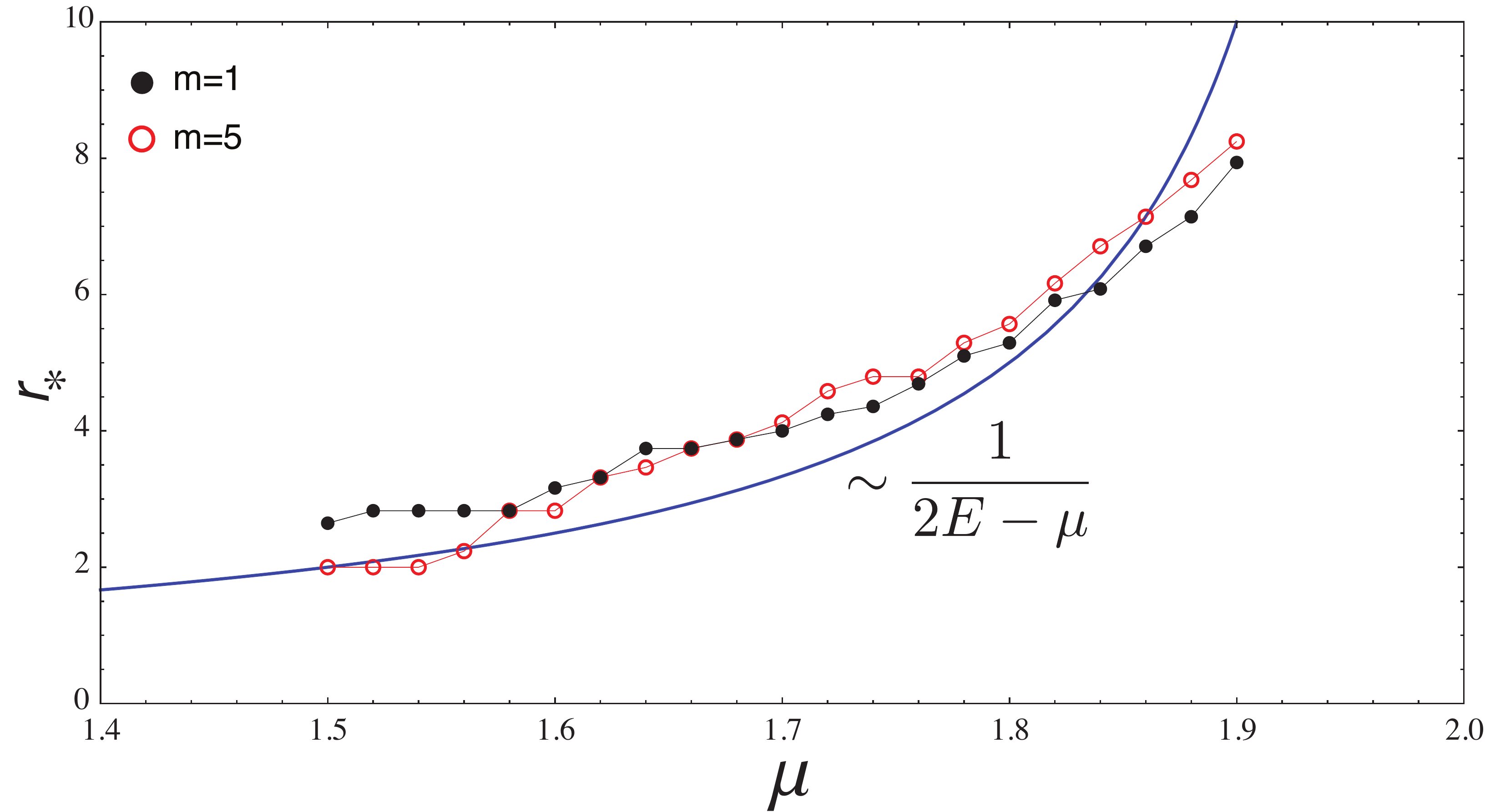}
\caption{\label{fig:rstarMu} Radius of the critical nucleation seed $r_*$ (here defined in units of tile length) as a function of $\mu$ extracted from data shown in Fig.~S\ref{fig:diss}. The data roughly agrees with the scaling predicted from nucleation theory (Eqn.~\eqref{eq:rstar}).}
\end{figure*}

\begin{figure*}[h!]
\centering
\includegraphics[width=0.95\linewidth]{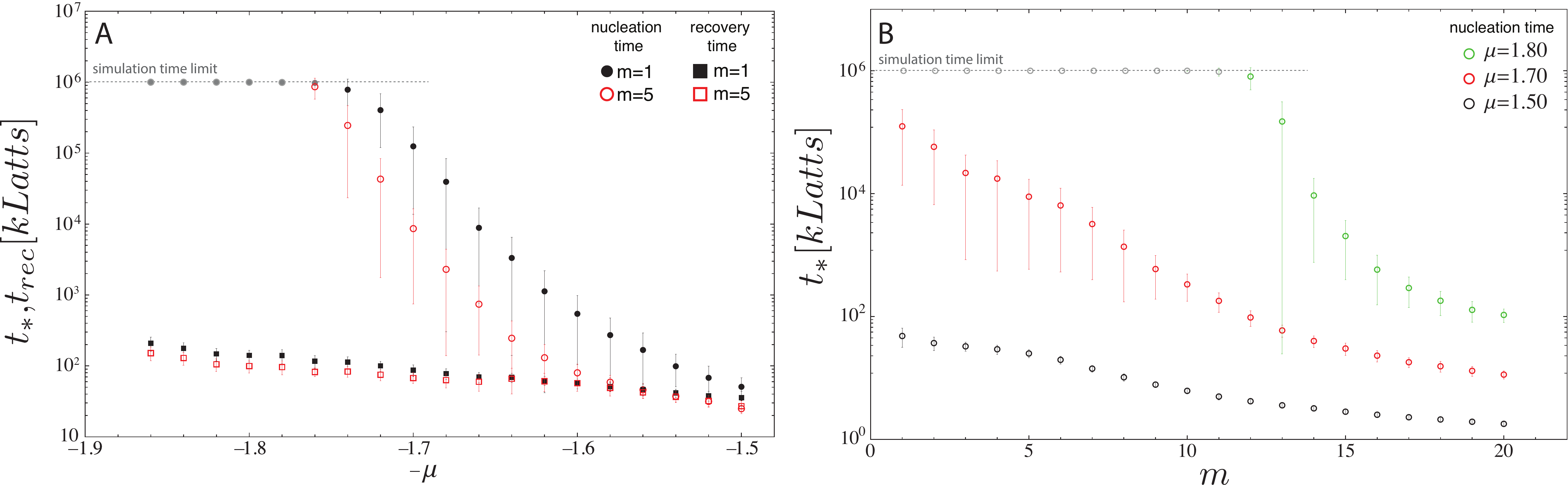}
\caption{\label{fig:tstarMum} A) Nucleation time $t_*$ and recovery time $t_{\text{rec}}$, in units of $kLatts(=10^3\text{ lattice sweeps})$, as functions of chemical potential $-\mu$, for $m=1\ \text{and}\ 5$ stored structures. The $t_{\text{rec}}$ does not significantly vary with $-\mu$, whereas $t_*$ shows a dramatic increase with $-\mu$. Each point is an ensemble average result over $100$ different runs. B) The $t_*$ vs. the number of stored structures $m$, for three different values of chemical potential $\mu$.}
\end{figure*}

\begin{figure*}[h!]
\centering
\includegraphics[width=0.5\linewidth]{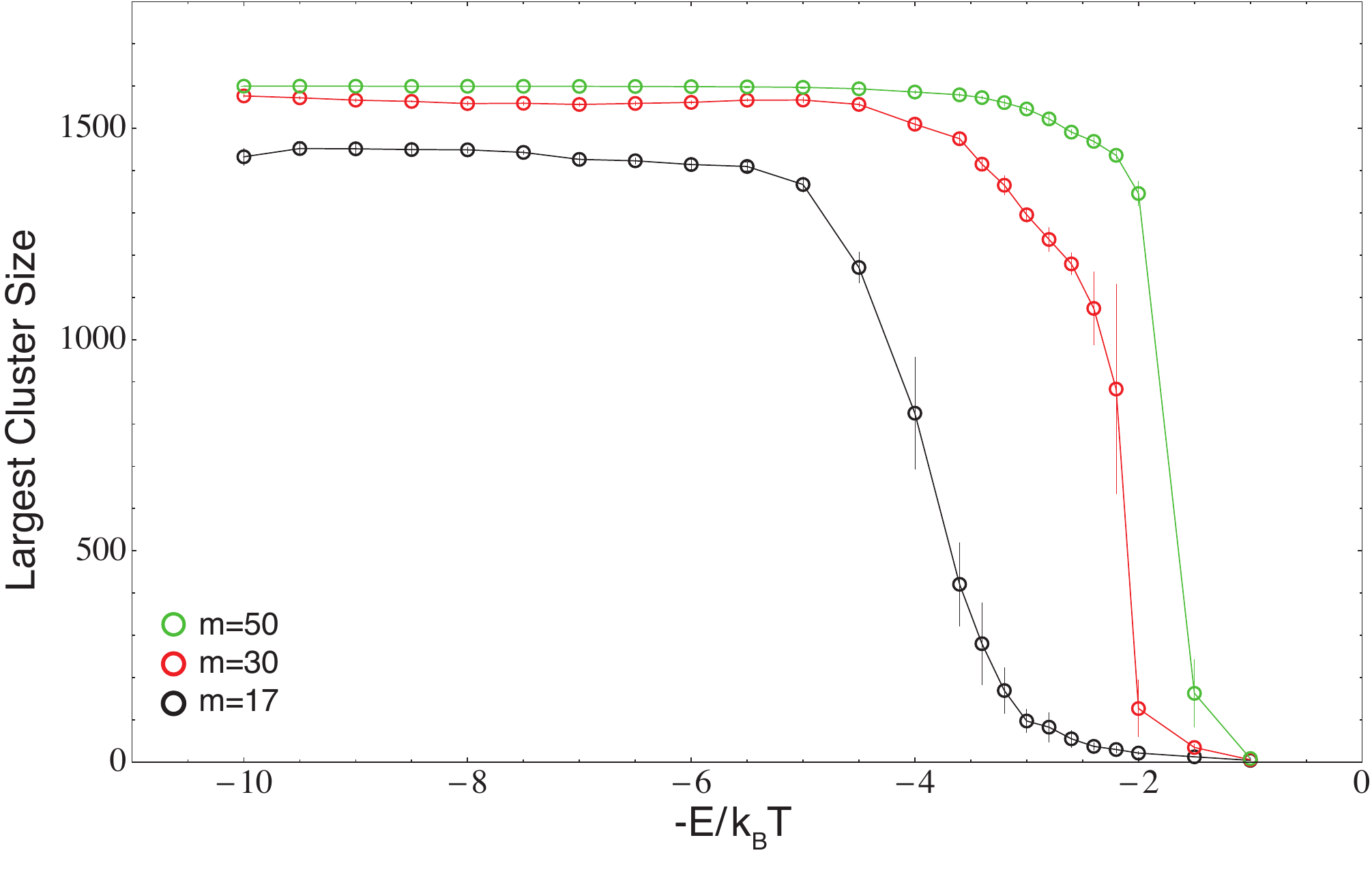}
\caption{\label{fig:transitionT} The finite temperature transition between regimes III and IV in Fig.~2, quantified by temperature dependence of the size of the biggest assembled structure. At the end of each simulation we identified the ``biggest structure'' as the largest connected area of tiles that are bonded by specific interactions. At each temperature its size was averaged over $10$ independent runs. At temperatures above the transition (regime IV), the system was also mostly filled with tiles, however, there were hardly any specific interactions between them, i.e., the state was a homogeneous solution of fluctuating components.}
\end{figure*}

\begin{figure*}[ht!]
\centering
\includegraphics[width=0.8\linewidth]{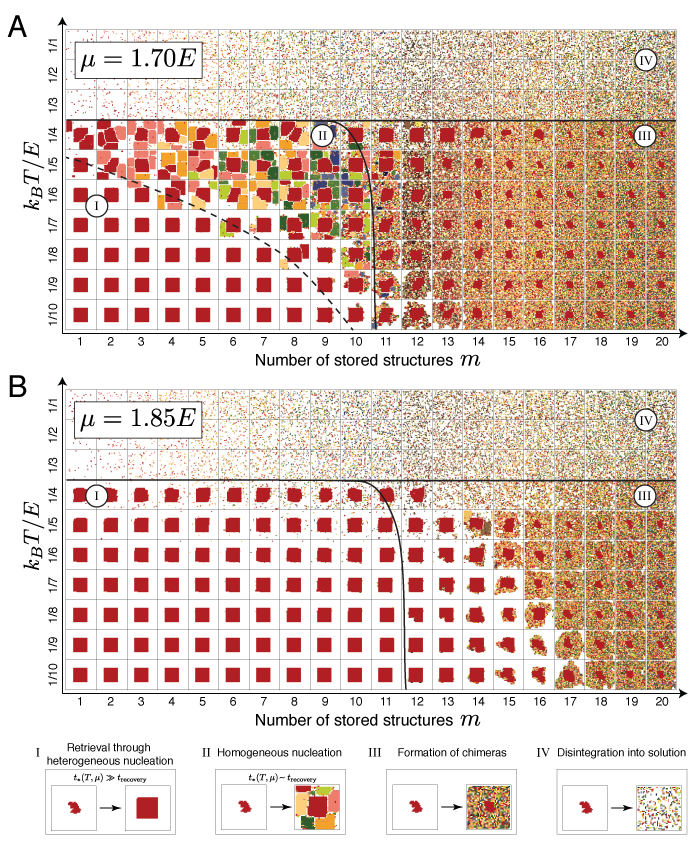}
\caption{\label{fig:regimesmu} The various regimes observed in Monte Carlo simulations as function of number of stored structures $m$ and temperature $k_BT/E$. These simulations were run for a fixed length of time, $2*10^6$ lattice sweeps. In comparison to Fig.~2 of main text the only difference is in the value of chemical potential $\mu$, which is equal for all species. A) $\mu=1.70E$: The lower value of the chemical potential influences the stability of the multifarious assembly mixture by reducing its lifetime $t_*$. Consequently the extent of the retrieval regime $I$ is suppressed by the homogeneous nucleation regime II. B) $\mu=1.85E$: For this value of the chemical potential, regime II is completely suppressed. This value of $\mu$ was used in Fig.~4 of the main text.}
\end{figure*}

\end{document}